\documentclass[fleqn,usenatbib]{mnras}

\usepackage{newtxtext,newtxmath}
\usepackage[T1]{fontenc}
\usepackage{ae,aecompl}
\usepackage{graphicx}	
\usepackage{amsmath}	
\usepackage{amssymb}	
\usepackage{booktabs}
\usepackage[table,xcdraw]{xcolor}
\usepackage{newtxtext,newtxmath}
\usepackage[T1]{fontenc}
\usepackage{ae,aecompl}
\usepackage{marvosym}
\usepackage{hyperref}
\usepackage{soul}

\newcommand{\beq}[1]{\begin{equation}\label{#1}}
\newcommand{\eeq}{\end{equation}}
\newcommand{\sub}[1]{_{\rm #1}}
\newcommand{\beqn}{\begin{equation}}
\newcommand{\eeqn}{\end{equation}}
\newcommand{\Op}{\Omega\sub{p}}

\newcommand{\asynch}{a\sub{synch}}

\newcommand{\am}{a\sub{m}}
\newcommand{\Mm}{M\sub{m}}
\newcommand{\ap}{a\sub{p}}
\newcommand{\astop}{a\sub{\mathrm{stop}}}

\newcommand{\Rs}{R_\star}
\newcommand{\Rm}{R_\mathrm{m}}
\newcommand{\Mp}{M\sub{p}}
\newcommand{\Rh}{R\sub{Hill}}
\newcommand{\Rp}{R\sub{p}}
\newcommand{\Pp}{P\sub{p}}
\newcommand{\Pm}{P\sub{m}}

\newcommand{\Der}{\mathrm{d}}

\newcommand{\Tors}{\tau\sub{p-*}}
\newcommand{\Torp}{\tau\sub{p-m}}

\newcommand{\Kpp}{k\sub{2}}
\newcommand{\Qp}{Q}

\newcommand{\Ip}{I\sub{p}}

\newcommand{\Lm}{L\sub{m}}
\newcommand{\Mmoon}{M\sub{m}}
\newcommand{\amoon}{a\sub{m}}
\newcommand{\nmm}{n\sub{m}}
\def\sgn{\mathrm{sgn}\,}
\newcommand{\Mstar}{M\sub{*}}
\newcommand{\apos}{a\sub{p}}
\newcommand{\npp}{n\sub{p}}

\newcommand{\slice}{kronian}

\definecolor{mycolor}{RGB}{150,1,255}
\definecolor{mycolorm}{RGB}{1,120,120}

\title[Can exoplanets preserve detectable moons?]{Can close-in giant exoplanets preserve detectable moons?}

\author[Sucerquia et al.]{\parbox{\textwidth}{
Mario Sucerquia$^{1,2}$\thanks{E-mail: \href{mailto:mario.sucerquia@uv.cl}{mario.sucerquia@uv.cl}},
Vanesa Ram{\'i}rez$^2$,
Jaime A. Alvarado-Montes$^{3,4}$\\
and Jorge I. Zuluaga$^2$.
}\vspace{0.4cm} \\
$^{1}$ N\'ucleo Milenio Formaci\'on Planetaria - NPF, Universidad de Valpara\'iso, Av. Gran Breta\~na 1111, Valpara\'iso, Chile \\
$^{2}$ SEAP,  
Instituto de F\'{\i}sica - FCEN, Universidad de Antioquia
Calle 70 No. 52-21, Medell\'in, Colombia.\\
$^{3}$ Centre for Astronomy, Astrophysics, and Astrophotonics, Macquarie University - Sydney, NSW 2109, Australia\\
$^{4}$ Department of Physics \& Astronomy, Macquarie University - Sydney, NSW 2109, Australia.}

\date{Accepted XXX. Received YYY; in original form ZZZ}
\pubyear{2019}

\begin{document}
\label{firstpage}
\pagerange{\pageref{firstpage}--\pageref{lastpage}}
\maketitle

\begin{abstract}
Exoplanet discoveries have motivated numerous efforts to find unseen populations of exomoons, yet they have been unsuccessful. A plausible explanation is that most discovered planets are located on close-in orbits, which would make their moons prone to tidal evolution and orbital detachment. In recent models of tidally-driven migration of exomoons, evolving planets might prevent what was considered their most plausible fate (i.e. colliding against their host planet), favouring scenarios where moons are pushed away and reach what we define as the \textit{satellite tidal orbital parking} distance ($\astop$), which is often within the critical limit for unstable orbits and depends mainly on the system's initial conditions: mass-ratio, semi-major axes, and rotational rates. 
By using semi-analytical calculations and numerical simulations, we calculate $\astop$ for different initial system parameters and constrain the transit detectability of exomoons around close-in planets. We found that systems with $\Mm/\Mp\geq 10^{-4}$, which are less likely to form, are also stable and detectable with present facilities (e.g. \textit{Kepler} and \textit{TESS}) through their direct and secondary effects in planet+moon transit, as they are massive, oversized, and migrate slowly. In contrast, systems with lower moon-to-planet mass ratios are \textit{ephemeral} and hardly detectable. Moreover, any detection, confirmation, and full characterisation would require both the short cadence capabilities of \textit{TESS} and high photometric sensitivity of ground-based observatories. Finally, despite the shortage of discovered long-period planets in currently available databases, the tidal migration model adopted in this work supports the idea that they are more likely to host the first detectable exomoon.

\end{abstract}



\section{Introduction}
\label{sec:intro}

The discovery of the first exomoon is a goal yet to be achieved. Using {\it Hubble} data, \citet{Teachey2018b} (hereafter TK18) announced recently the detection of a third anomalous transit of \textit{Kepler-1625b}. Along with other two transits detected in \textit{Kepler} data \citep*{Teachey2018}, this might hint the existence of an unconventional Neptune-sized exomoon.  The size of the object, temporarily designated as \textit{Kepler-1625b I},  is so large that we should better call it a `double planet'. However, this exomoon candidate is subject to debate as its signal has not been detected in subsequent studies. After TK18,  the first to investigate the combined \textit{Kepler} and \textit{Hubble} data into one dataset were \citet*{Heller2019}, who came up with alternative explanations for the exomoon interpretation. Also, by analysing the single \textit{Hubble} transit in TK18, \cite*{Kreidberg2019} have reported that the exomoon signal was not found. But if upcoming studies confirm it, this would be a remarkable discovery like that of the first exoplanet more than 20 years ago: 51 Pegasi b \citep{Mayor1995}. Still, future research is needed to determine whether the first exomoon will resemble the moons around the giant planets in the Solar System.

Direct observations of exomoons are also feasible by using IR wavelengths, at least at their most early stages of formation. \cite{Perez2019} has established observational limits on protolunar disc masses by means of direct images of exoplanets from observations with ALMA. In fact, the complex of radiotelescopes of ALMA has recently announced the first-ever detection of the early stages of exomoon formation around the giant exoplanet PDS~70c \citep{Isella2019}. This last work has determined that PDS~70c hosts a circumplanetary disc capable of forming moons, and comparing new ALMA observations with previous optical and infrared VLT observations, a disc-planet mass ratio of 10$^{-5}$\textendash10$^{-4}$ has been reported \citep{Muller2018}.

When a moon orbits a planet and this planet transits in front of its host star, two moon-induced simultaneous effects might arise in the system which are known as Transit Timing Variations (TTV) and Transit Duration Variations (TDV). Both of them can be detected at the same time \citep{Heller2016}, and require the transit of the planet but not necessarily the transit of the orbiting moon. The most comprehensive and self-consistent technique to discover and study exomoons is to perform an extensive photodynamical modelling, which besides including each of the aforementioned photometric effects in one framework, also integrates the celestial mechanics of a nested Keplerian star-planet-moon system \citep{Kipping2011,Rodenbeck2018}. There are many other methods that can be used to search for exomoons and characterize them (for a thorough and recent review please refer to \citealt{Heller2018}); however, and for the sake of simplicity, we will restrict here to a simple combination of transit light curve analysis and (barycentric) TTV+TDV.

Besides \textit{Kepler 1625b I}, the absence of other transits with moon-like signatures in the huge photometric  database of \textit{Kepler} \citep{Kipping2012,Kipping2013ApJ,Kipping2014ApJ}, favours the idea that exomoons might be `ephemeral objects', at least within the range of masses and orbital distances for the planets detected so far. The shortage of exomoon signals might be the aftermath of a long-term orbital instability of regular satellites (as we will suggest here). If this is the case, the detection of the first exomoon will probably shed light on the currently accepted models of formation, evolution and architecture of planetary systems. Additionally, it is worth noticing that most giant planets discovered so far lie within or close the habitable zone of their parent stars, so the discovery of any exomoon around those planets would also represent a significant step forward in the search of extraterrestrial habitable environments (see e.g. \citealt{Heller2013b,Heller2013,Heller2014}).

Several authors have proposed different mechanisms to explain the depletion of moons on relatively short time-scales around close-in planets. Some of them include: secular and resonant perturbations \citep{Barnes2002,Spalding2016}; planet scattering \citep{Hong2018}; and tidal detachment \citep{Sucerquia2019,Martinez2019}. 

From current formation models of giant planets and regular satellites, it is well-known that the initial physical and orbital properties of their moons such as mass $\Mmoon$, radius $r_\mathrm{s}$, and  semi-major axis $\amoon$ are not randomly distributed but closely related to the properties of the planet \citep{Canup2006,Ogihara2012,Heller2015,Cilibrasi2018}. For instance, it has been shown that the total mass of in-situ formed satellites is restricted to $10^{-4}\;M_{\rm p}$ (\citealt{Canup2006}, with $M_{\rm p}$ the planet's mass). Recent simulations of the formation of Galilean-like systems have revealed that satellite systems with masses from $10^{-3}$ to $10^{-2}\;M_{\rm p}$ could also exist \citep{Cilibrasi2018}.

If putative exomoons survive planetary migration (see \citealt{Namouni2010} and references therein), their tidal-induced orbital decay at the final planet's position will determine their final fate \citep*{Barnes2002,Alvarado2017}. Initial tidal models have suggested that in many cases, exomoons migrate outwards and survive temporarily before being pulled inwards at later stages \citep{Barnes2002}. Other moons, however, might exchange enough angular momentum via tidal interactions to be pushed farther away, becoming unstable and detaching from their parent planets. Many of these detached exomoons might collide with the planet, be absorbed by the star, or become new fully-fledged dwarf planets or planetary embryos \citep{Sucerquia2019}.

\citet{Alvarado2017} extended the initial orbital tidal-evolution models to include the  planetary radius contraction and interior structure evolution.  They found that in most cases the inward migration phase (if present) is suppressed.  If this physical scenario holds to be true, most regular exomoons would be parked at a `safe' asymptotic maximum distance, defined here as the \textit{satellite tidal orbital parking} ($\astop$), which depends on the initial $\amoon$, $\Mp$, planet's rotation and semi-major axis ($\apos$). This model could also have two interesting applications: Firstly, it helps us constrain the size of the signal produced by hypothetical satellites parked at $\astop$, and determine which systems will be detectable with present and future observational facilities. Secondly, by comparing the measured $\am$ with the predicted $\astop$ from future observations of exomoons, we could constrain the physical properties of the moon and the planet. Such models of evolving tides are also important for the orbital decay of planets (see e.g. \citealt{Alvarado2019}).

Here we explore hypothetical star-planet-moon systems, and start in Section \ref{sec:bio} establishing the regions where a moon would remain stable on its orbit during long time-scales. We then study in Section \ref{sec:prop} the distribution of $\astop$ for different combinations of parameters, and create therein synthetic exomoon populations around planets already discovered. The resulting effects of these moons on TTV and TDV signals are computed in Section \ref{sec:detection}, and the different implications of this work for past and future observations are discussed in Section \ref{sec:discussion}.

\begin{figure*}
{ 
\centering 
\includegraphics[scale=0.33]{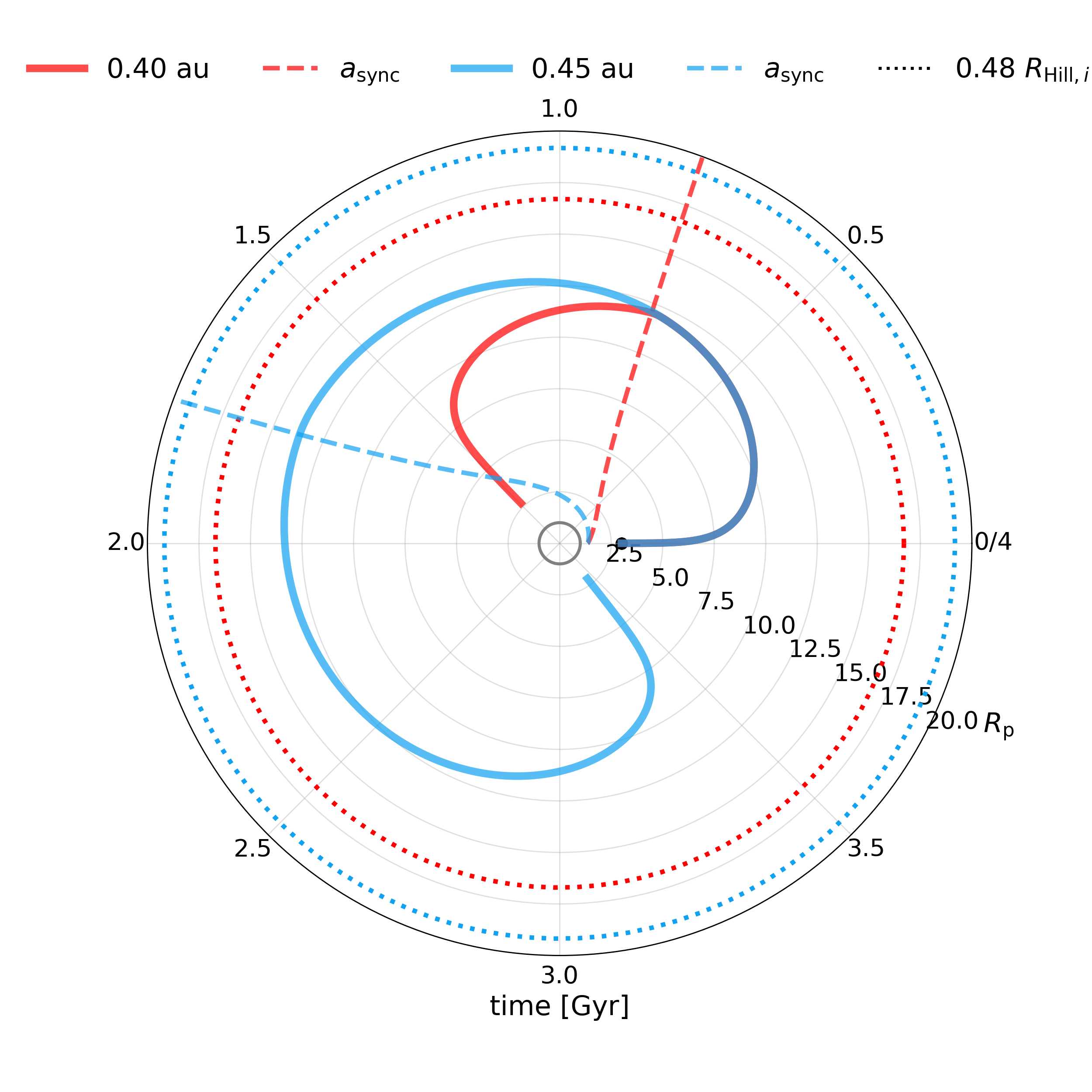}
\includegraphics[scale=0.33]{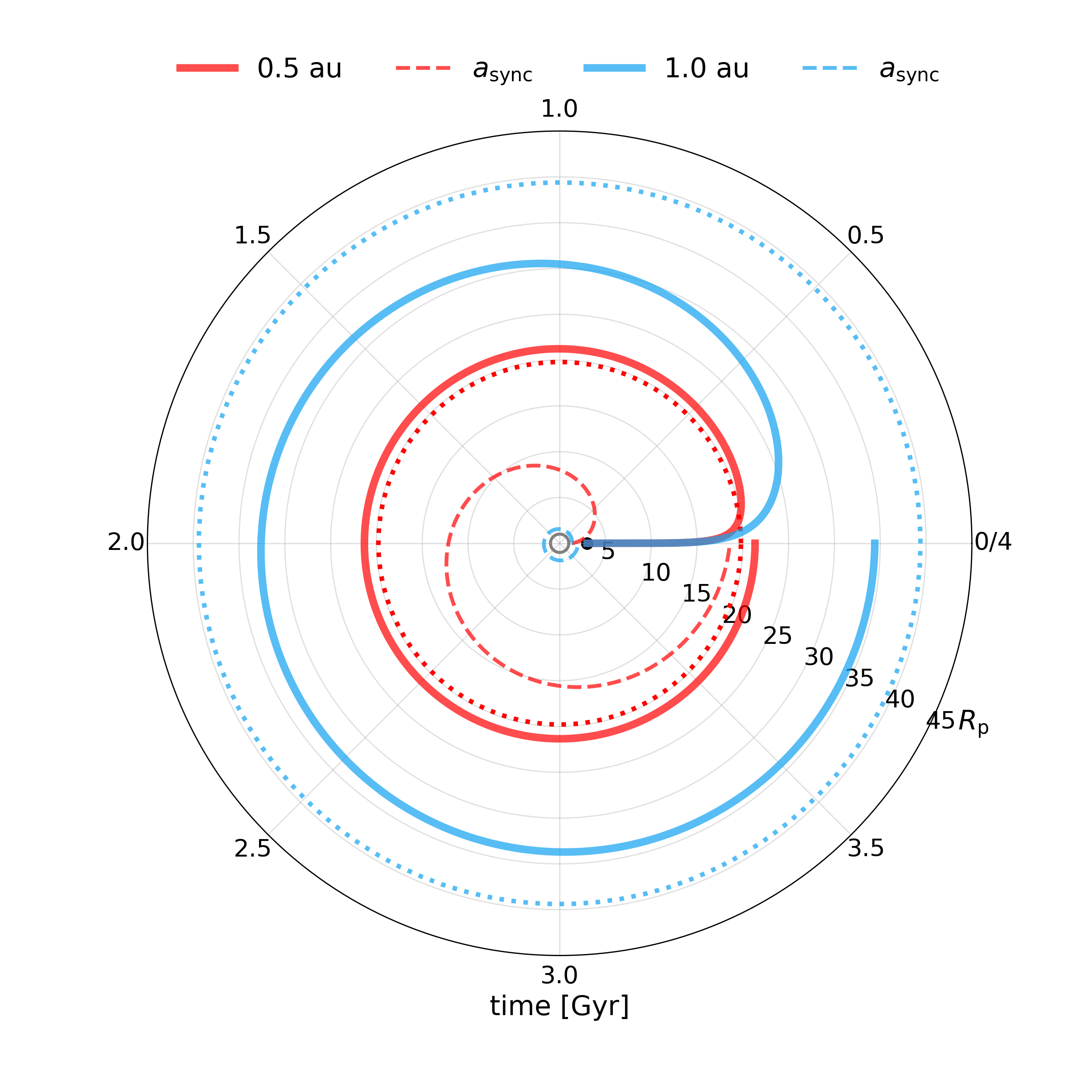}
\caption{Left-hand panel: orbital evolution according to the model of \citet{Barnes2002} of a moon starting in a planetocentric circular orbit at $3 \Rp$ . Simulations are ran for two different planetary orbits: 0.40 au and 0.45 au. dotted lines indicate the limit where prograde orbits become unstable (i.e. the secondary Hill radius $0.48\,R\sub{Hill}$), and dashed lines denote the position of the synchronous radius ($\asynch$) where $\Op=\nmm$. Right-hand panel: same as before but using the model of \citet{Alvarado2017}. Due to differences in time-scales, we show for illustration purposes the orbital evolution of an exomoon around a planet located at 0.5 and 1.0 au. In both plots time increases counterclockwise and ranges from 0 to $\sim$4.5 Gyr.} 
\label{fig:aevol}
}
\end{figure*}


\section{Tidally-driven orbital evolution}
\label{sec:bio}

Tidal interactions in a star-planet-moon system rise a bulge of complex geometry on the planet, which induces a transfer of angular momentum governed by two basic equations,

\beq{eq:difrot}
\frac{\Der\Op}{\Der t}\approx \frac{\Tors+\Torp}{\Ip}.
\eeq

\beq{eq:ratemoon}
\frac{\Der \Lm}{\Der t} = -\Torp,
\eeq

\noindent where $\Op$ and $\Ip$ are the planet's rotational rate and moment of inertia, respectively; $\Lm$ is the orbital angular momentum of the moon defined as $\Lm=\Mmoon\amoon^{2}\nmm$, with $\nmm$ the moon mean motion. From a constant time lag model \citep{Murray2000}, the tidal torques produced on the planet's bulge by the star ($\Tors$) and the moon ($\Torp$) are given by Eqs. (11) and (12) in \citet{Alvarado2017}.  After replacing the torque expressions into equations (\ref{eq:difrot}) and (\ref{eq:ratemoon}), we obtain the evolution equations for $\Op$ and $\nmm$:

\beq{eq:dOdt}
\begin{split}
\frac{\Der\Op}{\Der t}=-\frac{3}{2}\frac{\Kpp}{\Qp}\frac{\Rp^{3}}{\kappa^{2}G}
\left[
\frac{(G\Mstar)^{2}}{\Mp\apos^{6}}\,\sgn(\Op-\npp)+\right.\\
\left.\frac{\Mmoon^{2}}{\Mp^{3}}\nmm^{4}\,\sgn(\Op-\nmm) \right].
\end{split}
\eeq

\beq{eq:dndt}
\frac{\Der \nmm}{\Der t}=-\frac{9}{2}\frac{\Kpp}{\Qp}\frac{\Mmoon\Rp^{5}}{ G^{5/3}\Mp^{8/3}}\nmm^{16/3}\sgn(\Op-\nmm).
\eeq
which can be connected to the moon semi-major axis ($\amoon$) via Kepler's third law as $\amoon=(G\Mp/\nmm^2)^{1/3}$.

In the above equations $k_2$ and $Q$ are the planet's \textit{Love number} and tidal quality factor. If these equations are solved assuming that the physical properties of the planet ($\Rp$ and $k_\mathrm{2}/Q$) remain constant, the resulting evolution of $\amoon$ will follow a behaviour as in the left-hand panel of Fig. \ref{fig:aevol}. As noted previously by \citet{Barnes2002}, if the moon has an initial semi-major axis $\amoon$ where $\Op>\nmm$, the planet's tidal bulge exerts a positive torque on the moon which induces an outward migration. However, the dominant effect of the star is to spin down the planet's rotation (equation \ref{eq:difrot}), so when $\Op=\nmm$ at the synchronisation radius $\asynch$ (see Fig. \ref{fig:aevol}), the moon will start to migrate inwards because the sign of the torque (equation \ref{eq:dndt}) will be inverted.

\begin{figure*}
{ 
\centering 
\includegraphics[scale=0.37]{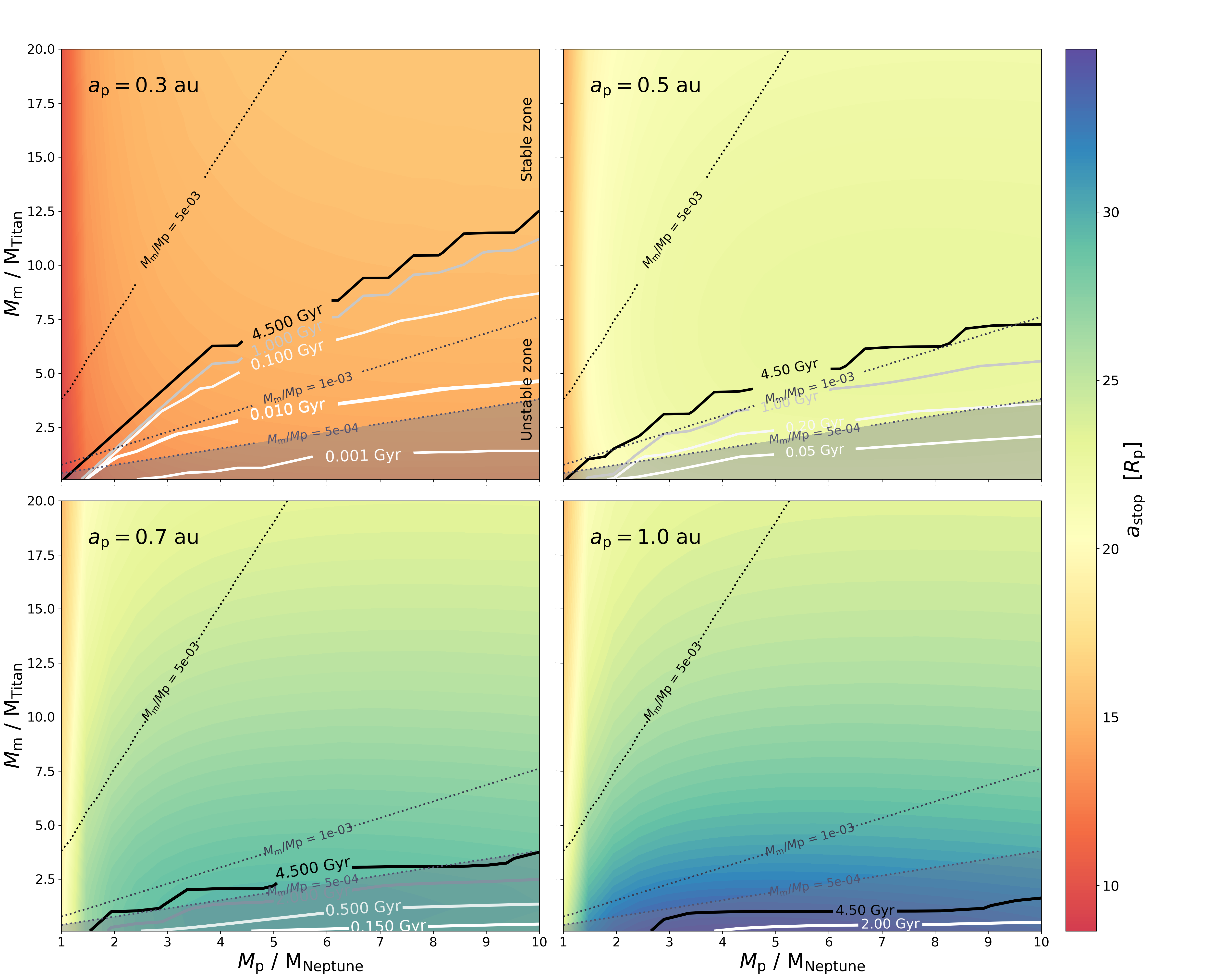}
\caption{Contours of the \textit{satellite tidal orbital parking} ($\astop$) distances for a tidally-migrated exomoon according to the model of \citet{Alvarado2017}.  Most of the remaining system parameters are in the \slice\, slice (see main text). Solid contour lines correspond to the time spent for a moon before reaching the stability limit of prograde orbits (i.e. $\sim0.48 \Rh$). Dotted contour lines correspond to the $\Mm/\Mp$ ratio for each system. 
The shaded region on the bottom of each plot correspond to systems having $\Mm/\Mp\lesssim10^{-4}$ which are the preferred value for in-situ formation models \citep{Canup2006}.}
\label{fig:tidev}
}
\end{figure*}

The system's properties, especially the planet's semi-major axis (different colours in Fig. \ref{fig:aevol}), change the moon synchronisation time. In most cases, however, moons eventually collide with their planet (see the left-hand panel of Fig. \ref{fig:aevol}). A different fate for moons might be possible if we accept that $\Rp$ contracts, and if we model the temporal change of the planet's tidal properties $\Kpp$ and $Q$ \citep{Alvarado2017} the orbital evolution of the moon will be similar to the curves shown in the right-hand panel of Fig. \ref{fig:aevol}. From this model, the rapid evolution of the planet during the first hundreds of Myr induces a faster outward migration and places the moon at the \textit{safe} distance $\astop$.  In some cases (red dotted curve in the right-hand panel of the same Figure), $\astop$ is larger than the critical Hill radius and the orbit of the moon becomes unstable, but in other scenarios (blue dotted curve) the moon remains parked for a few Gyr at $\astop$.

\section{Synthetic population of exomoons}
\label{sec:prop}

In the orbital evolution model described in the previous section, moons might be ejected from planetocentric orbits after interchanging angular momentum with its evolving planet; the probability for this to happen decreases as $\apos$ increases (see the right-hand panel in Fig. \ref{fig:aevol}). Although $\apos$ is the overarching parameter driving moon migration, other parameters in the system can also affect moon fate and migration rate.  This includes (but is not necessarily restricted to) $\Mmoon$, $\amoon$, planet's composition and initial rotational period $P_\mathrm{p}$.

Exploring exhaustively a parameter space having more than 5 dimensions, is simply unfeasible.  Still, we can learn several interesting lessons even if we confine our exploration to a specific `slice' of the configuration space.  We will call this the \textit{\slice\, slice}\footnote{The name of the slice, which clearly refers to Saturn (\textit{Kronos}), comes from the fact that the fixed properties of the planets and the moons in our synthetic population (initial planetary rotation period, planetary interior structure parameters, moon composition after removing all water content, etc.) are similar to that of Saturn and an ice-free Titan.}. In this slice, all systems are located around a solar-mass star; we fixed the initial moon orbital radius to $\am=3.0\,\Rp$; the rotation period of the planet is assumed $P_\mathrm{p}\sim$11 h, which is\footnote{Although initial planetary rotation seems to be rather arbitrary, we use this value for two main reasons: (1) it is similar to the rotational period of Jupiter ($\sim$10 hours), Saturn ($\sim$11 hours), Uranus ($\sim$17 hours) and Neptune ($\sim$16 hours); and (2) the corresponding initial synchronous radius is between $1.5-2$ $\Rp$ which is close to the Roche radii.}, in all cases, larger than the break-up period $P\sub{break}\sim 2\pi\sqrt{G\Mp/\Rp^3}$; and the planetary interior structure parameters are similar to those of Saturn (see \citealt{Alvarado2017}) with $\alpha = R_\mathrm{c}/R_\mathrm{p}=0.219$, and $\beta=M_\mathrm{c}/M_\mathrm{p}=0.196$, where $R_\mathrm{c}$ and $M_\mathrm{c}$ are the radius and mass of the planet's core, respectively.  In all cases we assume that the planetary and moon orbits are coplanar and circular. With this in mind, the \slice\, slice has (at least) three free dimensions: moon mass $M_\mathrm{m}$, planetary mass $M_\mathrm{p}$ and planetary semi-major axis $a_\mathrm{p}$.

Our synthetic population is comprised by systems with a discrete set of values of $a_\mathrm{p}=0.3,\,0.5,\,0.7$, and $1.0$ au, with moon masses ranging between 0.1 and 20 times the mass of Titan\footnote{For comparison the mass of Mars is approximately 5 times the mass of Titan.} (randomly generated assuming a flat distribution), and planetary masses ranging from 1 to 10 times the mass of Neptune\footnote{For comparison the mass of Saturn is approximately 5 times the mass of Neptune.} (randomly generated assuming a flat distribution). This will give us moon-planet mass ratios between $\Mm/\Mp\sim 10^{-5} - 10^{-2}$ which cover the whole range of formation models.

\begin{figure}
{ 
\centering 
\includegraphics[scale=0.34]{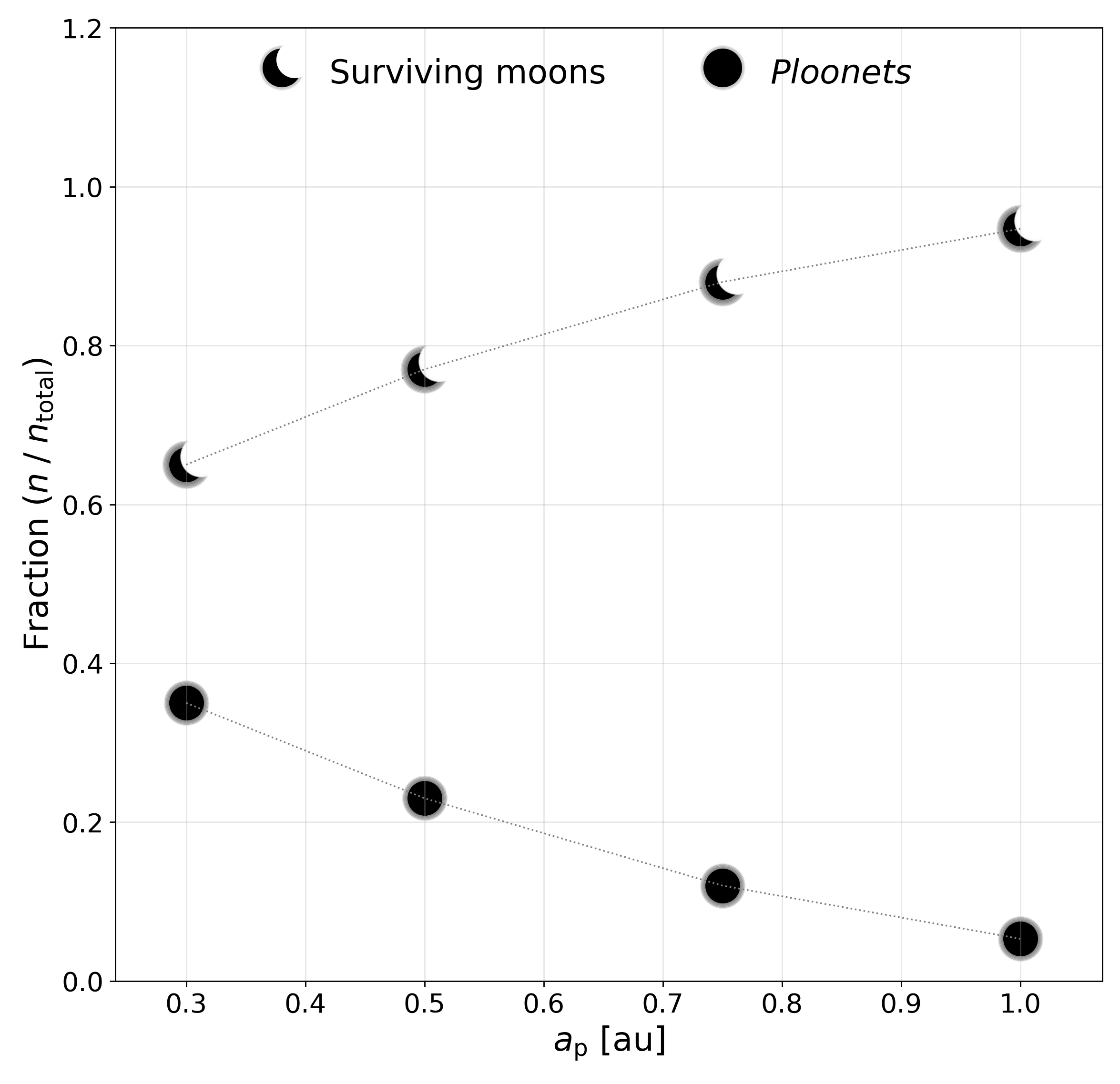}
\caption{Fraction of moons with stable orbits ({\it surviving}) and detached from the planet (ploonets), as a function of planetary semi-major axis $a_\mathrm{p}$ in the \slice\, slice of the parameter space (see main text). In all cases tidal evolution was integrated until 4.5~Gyr.
}
\label{fig:survival}
}
\end{figure}

Fig. \ref{fig:tidev} depicts contour plots of $\astop$ obtained after solving moon orbital migration using the \textit{realistic} model described in Section \ref{sec:bio}. 

As expected, the closer the planet the shorter the time-scale to reach the secondary Hill radius. Only those systems with $\Mm/\Mp$ above the thick solid  contour line (t$\sim4.5$~Gyr, the age of the Solar System) will survive and remain stable for relevant astrophysical times, and therefore will have larger chances to be detected.  We see that $\astop$ reaches several tens of $\Rp$ (at least in the \slice\, slice), reaching, in some cases, distances similar to that of the Earth-Moon system, $\amoon/\Rp\approx60$ (which we know is tidally-driven) and that presumed for the candidate \textit{Kepler-1625b} where $\amoon/\Rp\approx40$. Such large separation may also point out to a tidally-driven orbital evolution for this system.

\begin{figure*}
{ 
\centering 
\includegraphics[scale=0.34]{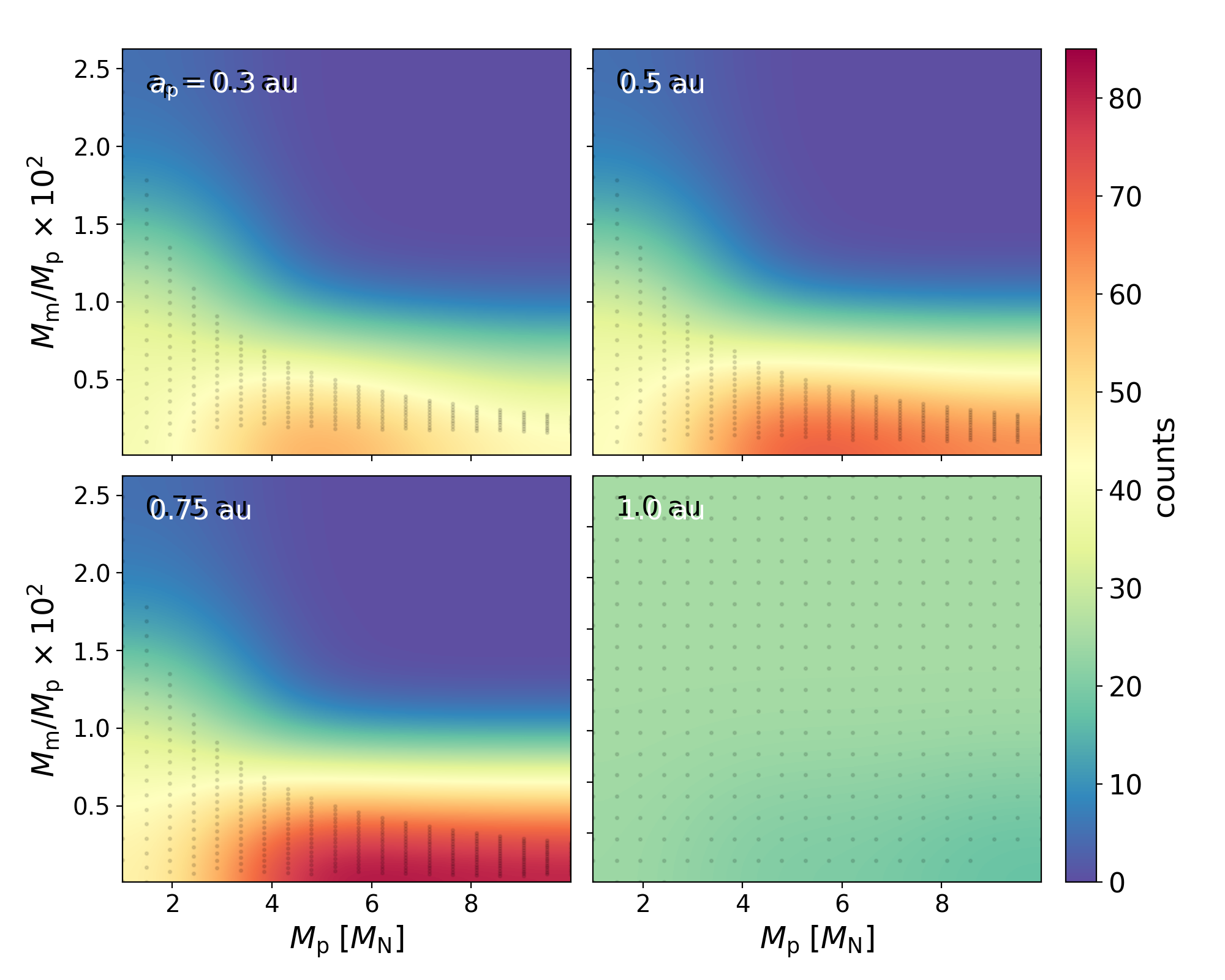}
\includegraphics[scale=0.34]{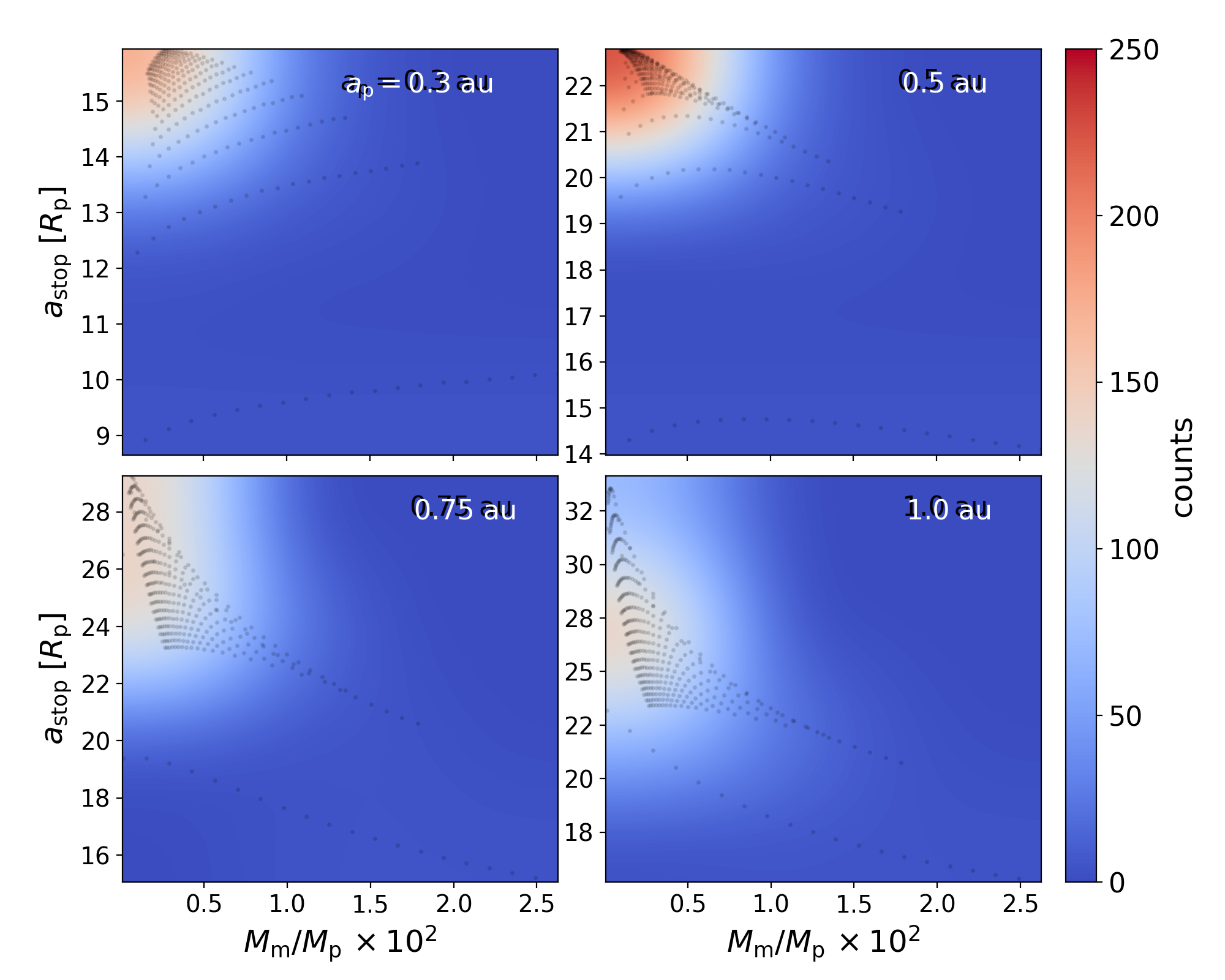}
\caption{Two-dimensional histograms of the occurrence rate of $\Mm/\Mp$ mass-ratios (left-hand panel) and maximum semi-major axis $a_\mathrm{stop}$ (right-hand panel) for the surviving moons of our synthetic sample (\slice\,slice), as a function of planetary mass and semi-major axis, $\Mp$, and $\ap$. Here $\Mp$ is given in units of Neptune's mass ($M_\mathrm{N}$).}
\label{fig:msas}
}
\end{figure*}

In Fig. \ref{fig:survival} we show the fraction of exomoons ($n$) attached ({\em surviving} tidal migration) and detached (mostly {\em ploonets}; \citealt{Sucerquia2019}) from its host planet after 4.5~Gyr evolution, respect to the total number of moons ($n\sub{total}$) and as a function of planetary semi-major axis. It can be seen that the chance for a satellite to survive increases with $\apos$. Low-mass moons (a dependency which is not shown in Fig. \ref{fig:survival}, but it is evident from the leftmost panel of Fig. \ref{fig:msas}) have a higher survival probability around more massive and more distant planets. Close-in and low-mass planets tend to preserve heavier moons (i.e. high $\Mm/\Mp$ ratios), and systems with low $\Mm/\Mp$ ratios tend to reach large $\astop$, increasing the vulnerability to dynamical instabilities (see the rightmost panel in Fig. \ref{fig:msas}).  Most of these trends are representative of the whole parameter space, even beyond the \slice\, slice assumed in this work.

The dynamical results obtained so far have been expressed in terms of the mass of the exomoons. However, since we are interested on their detectability via transits (see Section \ref{sec:transit}), we will need to assume a theoretical relationship connecting those masses to their corresponding radii. For this purpose, we assume that our synthetic moons are 100\% silicates (MgSiO$_3$) and use the correspondingly scaling-law for solid planetary bodies with this composition originally derived by  \citet{Fortney2007,Fortney2007b}, and recently improved by \citet{Awiphan2013}:

\beqn
R_\mathrm{m} = c_\mathrm{3} + c_\mathrm{2} \log \Mm + c_\mathrm{1} \; (\log \Mm)^2,
\label{eq:rm}
\eeqn
where $c_\mathrm{1}$=0.1567, $c_\mathrm{2}$=0.7275, and $c_\mathrm{3}$=1.1034. Here, both $\Mm$ and $\Rm$ are given in units of the Earth's mass and radius ($M_\oplus$ and $R_\oplus$, respectively). 

\section{Detectability}
\label{sec:detection}

The presence of a moon around a planet may cause (depending on the orbital parameters) an additional dip in the stellar light curve (transit), whose depth and duration depend on the size of the moon and its planetocentric semi-major axis, respectively. Despite a moon transit would have an extremely small depth, the smallest planets found so far are smaller than Mercury, e.g. \textit{Kepler-37b} \citep{Barclay_2013} which for comparison is $\sim1\%$ smaller than Saturn's moon Titan. Furthermore, if small exomoons are present in the \textit{Kepler} data, their orbital sampling effect would be potentially effective to find them \citep{Heller2014alone}.

The existence of exomoons may also be inferred from a small variation in the periodicity of planetary transits (TTV $\propto \Mm  \am$; \citealt{Sartoretti1999,Kipping09}), which are associated to changes in the planet's orbital position as a result of the gravitational pull of the moon\footnote{This effect can also be generated by the presence of other planets in the system, or even by previously detached exomoons or ploonets \citep{Sucerquia2019}.}. Exomoons may induce further observable changes in the planet's orbital speed that can be detected as deviations in the transit duration (TDV $\propto \Mm \am^{-1/2}$; \citealt{Kipping09,Kipping2009b}).

In the following sections we will  estimate the magnitude of all those effects for the synthetic population of exomoons around close-in planets studied in Section \ref{sec:prop}. These estimations will allow us to evaluate the detectability of tidally-evolved exomoons using current observational facilities.

\subsection{Eclipse effects}
\label{sec:transit}

\textit{Auxiliary transits} and \textit{mutual events} are produced by the combined effect of the planet and the moon when they transit in front of the host star. These so-called eclipse effects are useful to derive the size of the moon, and might be combined with TTVs and TDVs methods to get other bulk properties like density and composition.

The two types of eclipses are produced when the planet and moon apparent discs overlap (mutual events), and when they do not (auxiliary transits). The first case only occurs in edge-on systems aligned with the observer's line of view, so they are very difficult to perceive.

The depth of auxiliary transits for the moons in our simulations can be estimated by calculating the minimum stellar flux during the transit, $I_{*,min}$, for a planet with and without a moon. This happens when the planet and/or the moon discs are fully inside the stellar disc at the time of maximum eclipse, so $I_{*,min}$ can be written in terms of dimensionless quantities \citep{Seager03},

\beq{eq:imin}
I\sub{\star,min}= 1 - \left[ \left(\frac{\Rp}{\Rs} \right)^2 +\left(\frac{\Rm}{\Rs}\right)^2 \right] =  1 - (\delta_\mathrm{p} + \delta_\mathrm{m}).
\eeq
with $\delta_\mathrm{p}$ and $\delta_\mathrm{m}$ being the transit maximum depth produced by the planet and the moon, respectively. We found that exomoon radii range from 0.24 to 0.80 $R_\oplus$ for the synthetic population studied in Section \ref{sec:bio}, with large moons being the most frequent. Thus, if we isolate the contribution of the moon in equation (\ref{eq:imin}), $\delta_\mathrm{m}=(\Rm/\Rs)^2$ can be computed directly with equation (\ref{eq:rm}). 

\begin{figure}
{ 
\centering
\includegraphics[scale=0.4]{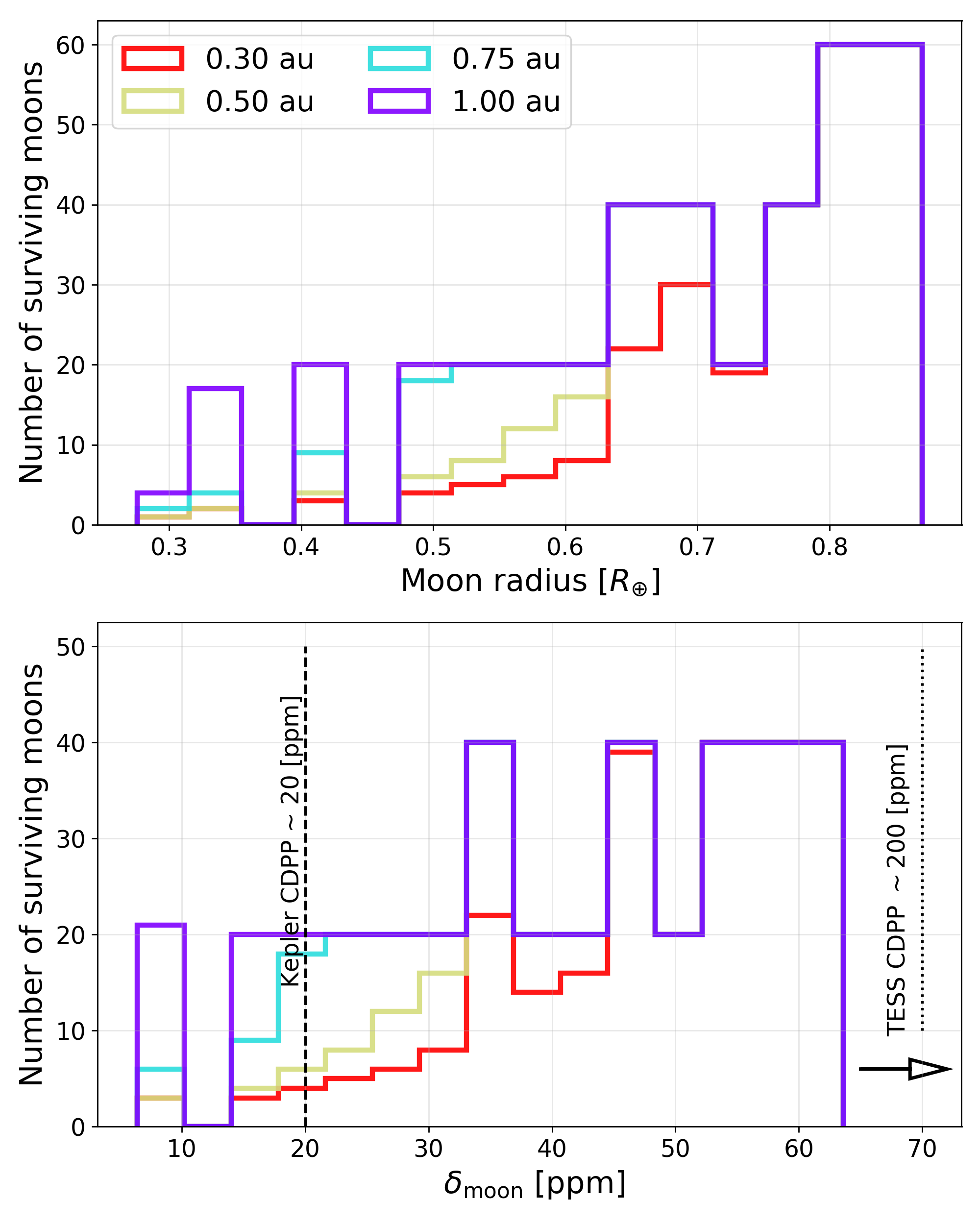}
\caption{Distribution of satellite sizes (upper panel) and transit depths (lower panel) for the surviving moons in our synthetic sample (\slice\, slice). Histogram colours indicate different values of the planetary semi-major axis $\ap$. Vertical lines in the lower panel indicates the value of the combined differential photometric precision (CDPP) for both Kepler and TESS data.  The CDPP of the later one is out of the range of the plot (arrow).}
\label{fig:delta}
}
\end{figure}

In Fig. \ref{fig:delta} we plot histograms (number of surviving exomoons) of two physical and transit key properties: the radius of the surviving exomoons (upper panel) and the exomoon maximum transit depth (lower panel). For the latter quantity we additionally show the so-called {\it combined differential photometric precision (CDPP)} for both \textit{Kepler} and \textit{TESS} observations.  In our case, CDPP measures the maximum photometric sensitivity required to detect variations in $I\sub{\star,min}$ induced by the presence of an exomoon (second term in \ref{eq:imin}).  Systems with exomoons producing transit depth variations to the left of the CDPP threshold could not be detected. 

Therefore, as revealed by Fig. \ref{fig:delta} the transit of most exomoons in the systems considered in this work can be potentially detected by Kepler. In contrast, TESS has not the required sensitivity to measure transit depth variations compatible with the mass-ranges and tidal evolution models assumed here.

\subsection{TTV and TDV effects}
\label{subs:TVs}

TTVs and TDVs are periodic signals that can be measured from systematic consecutive observations of a particular system. For any instrument to be able to detect these signals, the sensitivity must be high enough to spot the changes in the stellar brightness associated with the planetary transit, and have a cadence lower than the period of the signals.

The optimal TTVs occurs when orbits are circular and coplanar, for both the planet and moon. From the Earth's line of sight, and for edge-on orbits, the barycentric TTVs and their amplitude error ($\delta$) are given by

\beqn
TTV = \left[ \frac{\am \Mm \Pp }{2 \pi \ap \Mp}\right] \cos{f}
\label{eq:ttv}
\eeqn
and 
\beqn
\delta TTV =  \frac{\am \Mm \Pp }{\sqrt{2 \pi}\; \ap \Mp}, 
\label{eq:dttv}
\eeqn
respectively. In these equations, $\Pp$ is the orbital period of the planet and $f$ is the true anomaly of the lunar orbit.  

For optimal TDV, the best occurrence appears in systems whose orbits are circular and mutually inclined 90$^{\circ}$. In those cases, the barycentric TDV and its corresponding amplitude error are computed as follows,


\begin{figure}
{ 
\centering 
\includegraphics[scale=0.4]{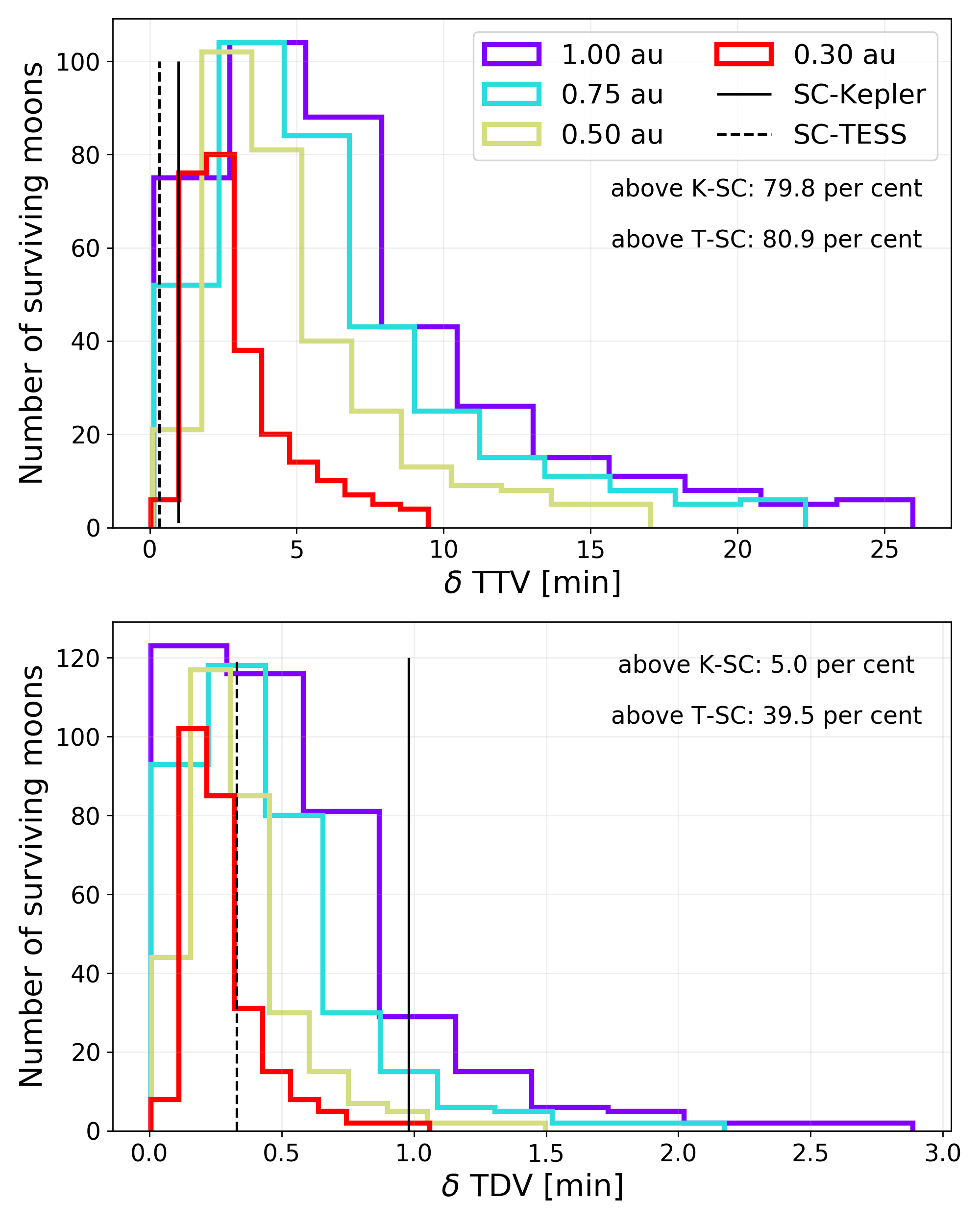}
\caption{Distribution of amplitude errors for the barycentric TTV ($\delta$TTV in upper panel) and barycentric TDV ($\delta$TDV in lower panel), expected for transits of the synthetic population (\slice\,slice). Vertical lines indicate the short-cadence limit for \textit{Kepler} (solid) and \textit{TESS} (dashed). In the legend we have included the fraction of surviving exomoons whose TTV and TDV signals might be detectable by \textit{Kepler} and \textit{TESS}.
}
\label{fig:txv}
}
\end{figure}

\beqn
TDV = \bar{\tau}\left[ \frac{\am \Mm \Pp }{ \ap \Mp \Pm}\right] \sin{f}
\label{eq:tdv}
\eeqn
and 
\beqn
\delta TDV =  \bar{\tau} \frac{\am \Mm \Pp }{\sqrt{2 \pi} \; \ap \Mp \Pm},
\label{eq:dtdv}
\eeqn
where $\bar{\tau}$ is the average duration of the transit and $\Pm$ the orbital period of the moon.

It is important to emphasize here that the analytical approximations in equations (\ref{eq:ttv}) and (\ref{eq:dtdv}) implicitly assume that the planet and the moon do not change their relative position during the transit (i.e. that moon orbital period is much larger than transit time). On the one hand, this assumption is reasonable since: (1) $a_\mathrm{stop}$ is typically large (10-40 $\Rp$, see Fig. \ref{fig:tidev}) and (2) planetary orbital mean motion is also relatively large for close-in planets. On the other hand, the transit signal of large exomoons will contaminate the light curve and hamper the measurement of the planetary TTV. However, since our aim here is just to estimate the magnitude of these effects instead of performing the actual analysis of the light curves for our synthetic transits, our conclusions will not be significant modified by the assumption that equations (\ref{eq:ttv}) and (\ref{eq:dttv}) provide a good estimation of the actual effect.

With the aim of obtaining the aforementioned variations for each system, in this section we use the outcomes of the simulations of moon migration described in Section \ref{sec:bio} as inputs for equations (\ref{eq:dttv}) and (\ref{eq:dtdv}). Particularly, we use the values computed for $\astop$ associated with each pair of $\Mp$ and $\Mm$, where moons still remain in stable orbits around the planet. This approach is different than that of starting from a given TDVs and obtaining from it the values of $\am$ and $\Mm$, as originally intended.

\begin{figure*}
{ 
\centering 
\includegraphics[scale=0.36]{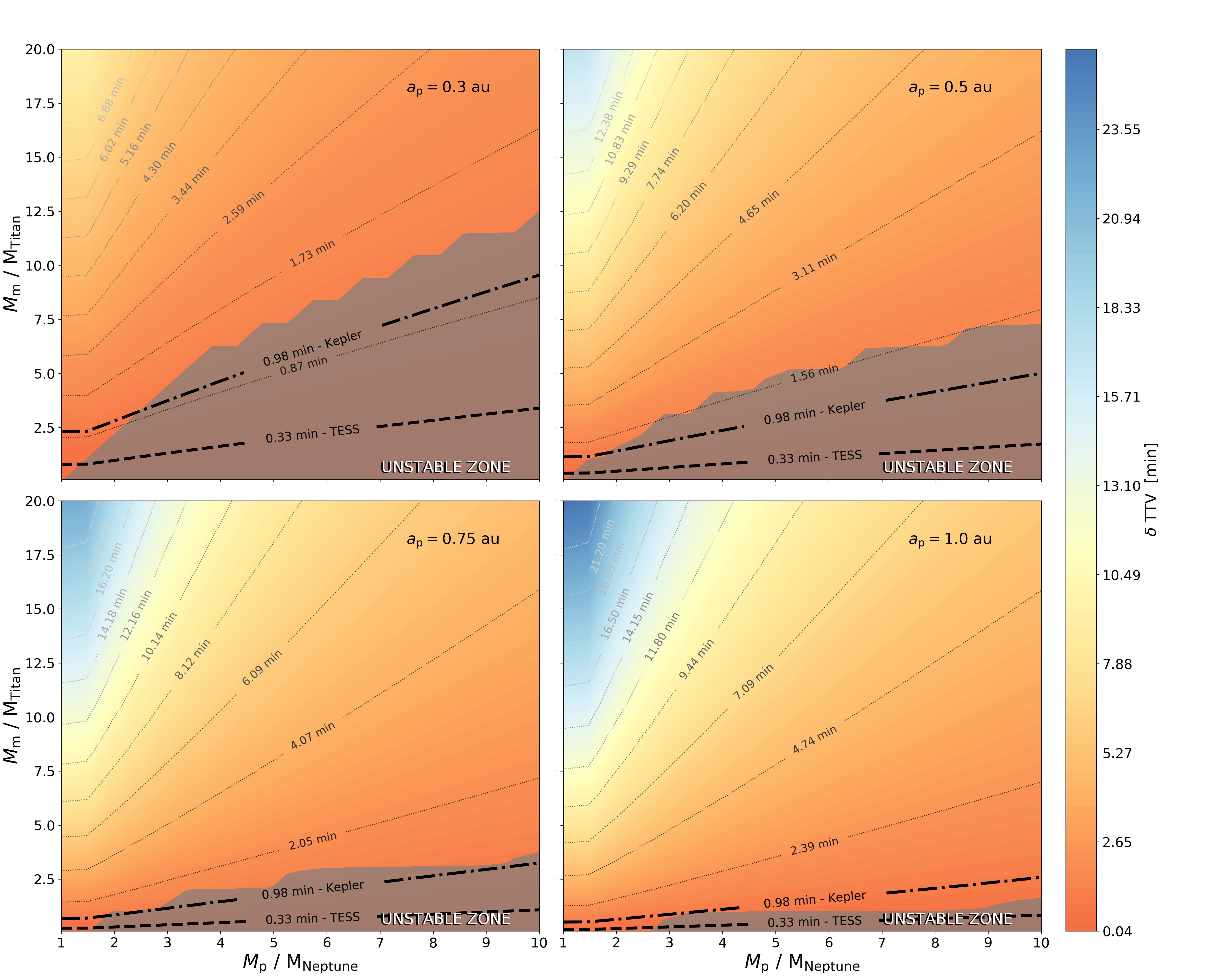}
\caption{Colour maps of the barycentric TTV signals for tidally-migrated exomoons at different planetary semi major axis (upper-right label in each plot). In all cases we are assuming conditions in the so-called \slice\, slice of the parameter space. Contour lines have been added for clarity., and as a reference we show
the contours corresponding to the short cadence limit of \textit{Kepler} (dash-dotted) and \textit{TESS} (dashed). The shaded areas in the bottom part of the contour plots correspond to system which after $\sim 4.5$ Gyr of tidal evolution are in potentially unstable orbits.}
\label{fig:TTVS}
}
\end{figure*}

Fig. \ref{fig:txv} shows the distribution of amplitude errors ($\delta$TTV and $\delta$TDV) for the surviving systems studied in this work. Small values of $\ap$ lead to sharp peaked distributions and produce lower signals as a result of their smaller Hill radii, while large $\ap$ produce flatter amplitude error distributions. This happens because at large distances low-mass moons have slow migration rates, therefore they are still drifting from their initial orbits. Also, more distant planets have a larger number of remaining exomoons and hence they dominate the total distribution of $\delta$TTV and $\delta$TDV.

To answer the question whether these synthetic systems are detectable, we  compare their TTV and TDV signal errors with the timing uncertainties at short cadences (SC) of \textit{Kepler} \citep{Gilliland2010,Ford2012} (solid vertical line) and \textit{TESS}\footnote{\url{https://tess.mit.edu/science/observations/}} (solid dashed line).  In addition, we compute and show in the figure legend, the fraction of detectable exomoons among all synthetic systems studied in this work.

\begin{figure*}
{ 
\centering 
\includegraphics[scale=0.36]{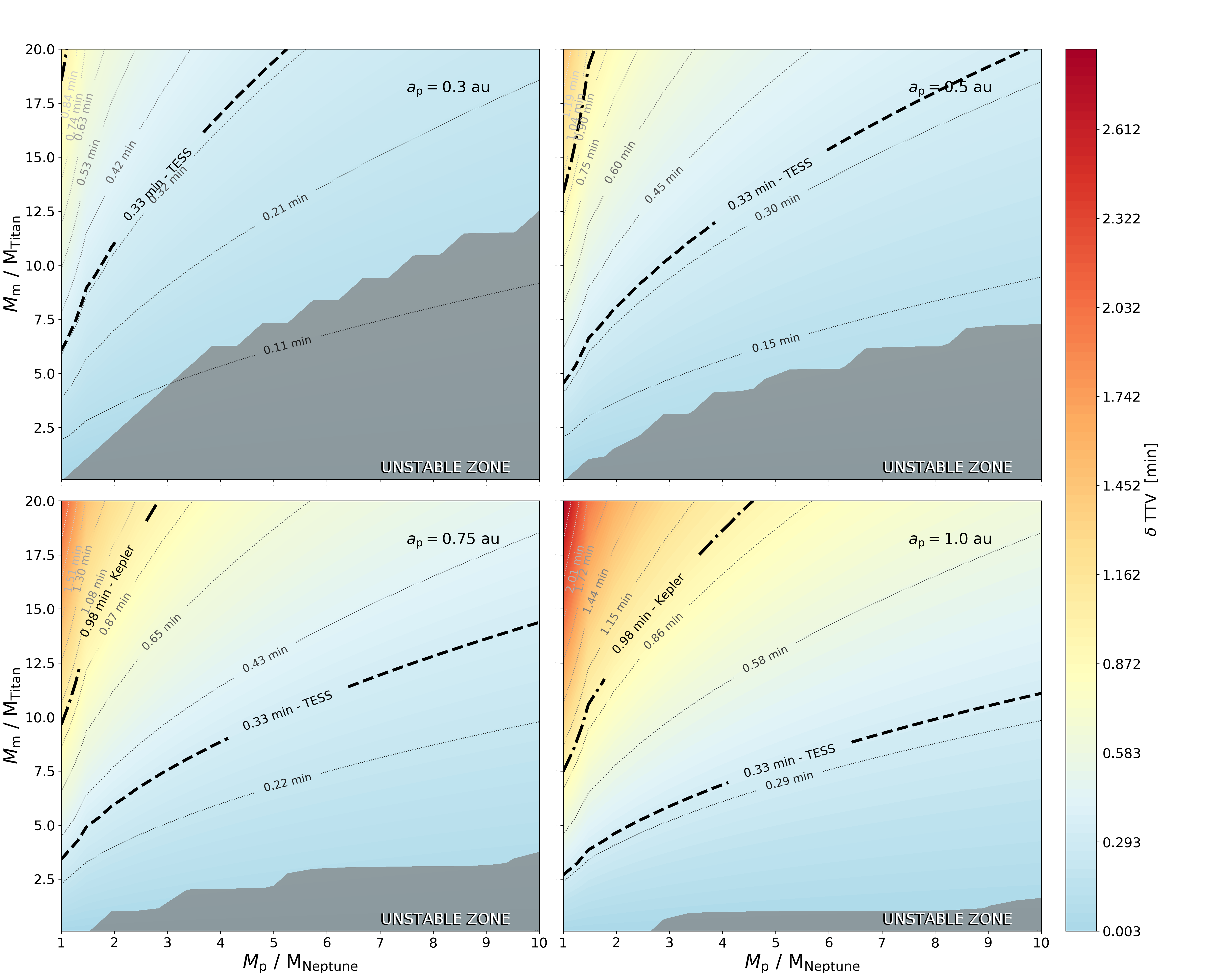}
\caption{
Same as \autoref{fig:txv} but for the case of barycentric TDV.
\label{fig:TDVS}
}
}
\end{figure*}

The actual values of the TTV and TDV (Fig. \ref{fig:TDVS}) for the complete set of synthetic systems considered in this work, are shown by the contour maps of Figs \ref{fig:TTVS} and \ref{fig:TDVS}. The overlapped contour lines show the amplitude errors for the associated signals (TTV or TDV). We have also shown (dashed and dash-dotted lines) the instrument limit at minimum cadence of \textit{TESS} and \textit{Kepler}, respectively.  In terms of detectability via TTV and TDV, systems made of a large moon orbiting a small planet are clearly favoured. However, despite the fact that systems hosting small moons are more dynamically unstable, a non-negligible fraction of them could be detected by \textit{TESS} and \textit{Kepler}. Measuring TDVs is far less promising, especially in the case of \textit{Kepler} data.  However, \textit{TESS} might still be capable of detecting transit variations of large surviving exomoons around giants in habitable-zone. 

A comparison of the shaded regions in Figs \ref{fig:tidev}, \ref{fig:TDVS}, and \ref{fig:TTVS} (unstable zone) reveals that very few systems obeying current formation models ($M_\mathrm{m}/M_\mathrm{p}\sim 10^{-4}$), have TTV signals detectable with \textit{TESS} and \textit{Kepler}, and that none of those systems would produce TDVs within the sensitivity limit of both instruments. Nevertheless, the number of exomoon detections can be larger for more distant planets.

The degeneracy of $\am$ and $\Mm$ in equations (\ref{eq:dttv}) and (\ref{eq:dtdv}) can be solved, if we measure simultaneously the TTV and TDV effects, to find the bulk physical properties of a moon.  However, as shown before, this would only be possible for system with large exomoons, which besides being unstable, are hardly detectable by \textit{TESS}.

\vspace{-0.5cm}

\section{Summary and Discussions}
\label{sec:discussion}

Classical theories of moon tidal evolution assume that a planet's bulk and rheological properties do not evolve while its moons undergo migratory processes \citep{Barnes2002}. The introduction of evolutionary models for the size and internal structure of a planet \citep{Alvarado2017} modifies, or even restricts the final stages of moon orbital migration. In consequence, this imposes bounds on the masses ($\Mp$ and $\Mm$) and planetary semi-major axes ($\ap$) for which planets are able to preserve their moons once their tidal migration ends. At short $\ap$ migration of moons is significant and occurs quickly, which can lead to outward ejection (see e.g. \citealt{Sucerquia2019}) and orbital instabilities.

In this work we have studied the possibility for moons orbiting close-in giant exoplanets to survive migration caused by stellar and planetary tides, and explored the parameter space to estimate the number of surviving satellite systems (Fig. \ref{fig:survival}). We then found distinct migration regimes for these exomoons to succeed and settle to a fixed final position, so that their theoretical detectability by using current observational facilities could also be analysed.

Most cases studied here share common characteristics: planets are strongly affected by stellar tides, which slow down the planetary rotational rate; moons around light planets have slow migration rates due to a small transfer of angular momentum, and the orbits of light moons are not perturbed enough to reach far-off {\it satellite tidal orbital parking} ($\astop$) distances, as they do not produce significant tidal bulges on their planets. All of these effects are more notorious at short $\ap$.

As established in \cite{Kipping2009c}, TDV signals are \textit{exclusively} induced by moons. However, the synthetic population of moons explored in this work has shown that few of them are within the sensitivity limit of \textit{Kepler} and \textit{TESS} (see Figs \ref{fig:TTVS} and \ref{fig:TDVS}). Notwithstanding that the detection probability of these moons is low owing to their small size and short $\astop$, all of them would be detectable with \textit{Kepler} if we use {\it Transit Timing Variation} (TTV) signals, but their existence has to be confirmed with other methods since TTVs can also arise due to interactions with other planets in the system (even with previously detached exomoons or ploonets; \citealt{Sucerquia2019}).

Large satellites produce substantial tidal bulges on massive planets, enhancing thereby an efficient transfer of angular momentum while the planet's rotation decreases at a slow rate. With large planets having a large moment of inertia, moons are gradually pushed away towards distant positions where their orbital detachment might occur. However, very large satellites migrate slower because of their large orbital inertia. The regime of rapid migration is mainly composed by moons whose mass surpasses the expected values from formation models (i.e. $\Mm/\Mp \sim 10^{-4}$; \citealt{Canup2006}), which are more prone to be ejected from their orbits when planets are located at close-in $\ap$ where the stellar torques are stronger and the Hill sphere is smaller. Still, for these moons we can expect well-defined signals of {\it Transit Duration Variation} (TDV) and TTV (Fig. \ref{fig:txv}) within suitable detectability zones (see Figs \ref{fig:TTVS} and \ref{fig:TDVS}).

For all the explored scenarios we considered a system composed by a single moon orbiting an isolated planet. Multiple moons in a system can change moon migration and lead to diverse results. For instance, resonant configurations in the system might produce orbital chaos and trigger collisions among satellites. This in turn could spark the formation of planetary rings or boost the ejection of moons. Also, multiple satellites could remove TTV and TDV signals, and this would hinder even more the transit detection of the system. Complex scenarios like this might be circumvented by adopting supplementary methods, e.g. \textit{Orbital Sampling effects} \citep{Heller2014alone}, to increase the detectability of multiple exomoons. Unfortunately, moon sizes as those found and reported in Section \ref{sec:transit} will be really difficult to decode from real light curves.

In the scope of this work, we carried out a handful of simulations for the yet-to-be-confirmed exomoon \textit{Kepler 1625b I} \citep{Teachey2018}. A slow migration rate was found, and the most likely final position was found at $\sim 10 \Rp$. As a result, the system would exhibit TTVs of $\sim$ 0.5 and 4.5 min (i.e. within the sensitivity limit of \textit{Kepler}), but TDVs from 0.09 to 0.7 min which are not measurable by this facility. Although the procedure we followed predicts undetectable TDVs for \textit{Kepler 1625b I}, the scarce observations of this system have reported extremely large masses, so mutual tides in the planet-moon system should be considered to study the early moon orbital migration \citep{Piro2018} and a possible pull-down capture origin \citep{Hamers2018}.

Exomoon detectability linked to tidal evolution is an ongoing and growing research area. In this pursuit, we aimed at adopting a more realistic moon migration model to assess the optimal combination of physical and orbital parameters that might increase the detectability of extrasolar satellites. This would allow us to elucidate the dynamical architecture of planet-moon systems and establish some constraints to identify evidence of moons in secondary transit effects such as TTV, TDV, and auxiliary transits. In this regard, novel observational methods carried out with current facilities are necessary to solve the degeneracy of $\am$ presented between TTVs and TDVs; and implementing precise models for planetary mass-radius relationships, in turn, can further enhance the detection of these signals.

To conclude, the two largest data sources of exoplanet photometry (i.e. the space-based telescopes \textit{Kepler} and \textit{TESS}) have chosen analogous targets regarding stellar age, fairly similar to our Sun. However, the downside of these stars is that they are too evolved and most of the moons around their close-in giant planets could have been obliterated (be detached or collided with the planet or with other moons). If we want to direct an effective exomoon hunt, the idea of selecting young stars targets is tempting since moon migration is still undergoing. Unfortunately, young stars are objects with a very high photometric activity and, consequently, few planets have been confirmed around them. In this regard, the detection of their exomoons becomes a challenging task. The outcomes presented here, complementary to previous works, support the idea that long-period planets whose detection is much less frequent are more prone to harbour detectable exomoons.

As mentioned in Section \ref{sec:prop}, all the specific results presented in this work are restricted to the so-called \slice\, slice of the parameter space, namely a thin section of a N-dimensional space of all possible physical properties in star-planet-moon systems.  It was not our purpose to describe exhaustively the detectability of all configurations, but to find some well-recognized patterns in a suitable region of the parameter space.  What would happen beyond the \slice\, slice? If instead of a solar mass star we perform the same simulations around a less massive star, many of the results here will probably scale-down to smaller distances $\ap$, without significant modifications (see equation \ref{eq:difrot}) with the exception probably of the time-scales of tidal-induced migration, which is not a minor effect.  If the period of rotation of the planet was as large as the largest period among solar system giants ($\sim$16 hours), the initial semi-major axis of the exomoons should be also larger ($a_\mathrm{sync}\propto P_\mathrm{p}^{2/3}$).  This would change the migration timescales and probably improve the chances to detect large exomoons before they become unstable.  In most of these cases, however, many of the trends identified here will not change significantly and the conclusions of this work will not be substantially modified. 

The incessant search of life outside the Solar System has proposed that this might emerge and thrive in moons around giant planets located within circumstellar habitable zones (see e.g. \citealt{Martinez-Rod2019}). However, an important effect of this work will be readdressing carefully the fact whether fast migration of moons entails enough favourable conditions to make these moonish environments compatible with life.

\vspace{-0.6cm}
\section*{Acknowledgements} 
The authors are really grateful with the referee, Dr. Ren\'e Heller, whose comments and suggestions significantly improved this work. Mario Sucerquia acknowledges funding support from Iniciativa Cient\'ifica Milenio (`ICM') via N\'ucleo Milenio de Formaci\'on Planetaria and Colciencias, and Jaime A. Alvarado-Montes is funded by Macquarie University through the International Macquarie University Research Excellence Scholarship (`iMQRES'). Jorge I. Zuluaga is supported by Vicerrector\'ia de Docencia UdeA. This article was written while thousands of people march on the streets in Colombia and Chile, seeking education and social equity for everybody. Great thanks to all of them. Este art\'iculo fue escrito mientras miles de personas marchan en Colombia y en Chile, en la b\'usqueda de educaci\'on e igualdad social para todos. Un agradecimiento a todos ellos.

\bibliographystyle{mnras}
\bibliography{references} 

\bsp	
\label{lastpage}
\end{document}